\documentclass[aip, jcp, reprint, floatfix, amsmath, amssymb, citeautoscript]{revtex4-2}% draft twocolumn preprint, longbibliography,

\usepackage{graphics, graphicx, bm, natbib, color}

\usepackage{graphics, graphicx, bm, natbib, color, xcolor}
\usepackage{comment}
\usepackage{appendix}
\usepackage{mathtools}
\usepackage{booktabs}
\usepackage{newtxtext, newtxmath}
\usepackage{anyfontsize} % Allows LaTeX to scale the default fonts smoothly to any custom AIP size
\usepackage[capitalise]{cleveref}
\newcommand{\rem}[1]{}

\newcommand{\ud}{\mathrm{d}}
\newcommand{\diag}{\text{diag}}

\newcommand{\wm}[1]{\mathbf{\underline{ #1}}}

\begin{document}
 \author{Wijnand Broer}
\email{wbroer@gmail.com}
\affiliation{Zernike Institute for Advanced Materials, University of Groningen,  9747 AG Groningen, The Netherlands }
%Nijenborgh 4,
\author{Jure Dobnikar}
\email{jure.dobnikar@fmf.uni-lj.si}
\affiliation{Chinese Academy of Sciences Key Laboratory of Soft Matter Physics, Institute of Physics, Chinese Academy of Sciences, Beijing 100190, China }
% \affiliation{School of Physical Sciences, University of Chinese Academy of Sciences,Beijing 100049, China}
\affiliation{Wenzhou Institute, University of Chinese Academy of Sciences, Wenzhou, Zhejiang 325011, China}
\affiliation{Faculty of Mathematics and Physics, University of Ljubljana, 1000 Ljubljana, Slovenia}
% Jadranska ulica 19,

\title{Casimir force in a water nanolayer confined by graphene sheets}

\date{29 September 2026}

\begin{abstract}

Casimir forces, arising from quantum and thermal fluctuations, depend strongly on the electromagnetic susceptibilities of the interacting materials. Here, we theoretically investigate the Casimir force between two graphene sheets separated by a nanometer-thin layer of liquid water. 
We find that deviations from the bulk case can be as large as 24 \%, mostly due to the static susceptibility components. This effect does not decrease monotonically as a function of the layer thickness: it increases up to a maximum at 70 nm and it slowly declines to the bulk limit only at 1 micron. This is due to two competing effects: on one hand, the anisotropy decreases with the layer thickness; on the other hand, the relative contribution of the most anisotropic, static susceptibility components increases.

%at thicknesses up to 3.4 nm, the confinement effect on the Casimir force is up to 20 \% compared to the bulk case. At larger thicknesses, the bulk model underestimates the Casimir force by up to 24 \% due to the interface effect. This effect does not decrease monotonically as a function of the layer thickness: it increases up to a maximum at 70 nm and it slowly declines to the bulk limit only at 1 micron. 
%We separately assess the effect of nanoscale confinement on  the electronic response of graphene, which has no effect on the Casimir force.  

\end{abstract}
\maketitle
 
 \section{Introduction}
  The Casimir effect \cite{Casimir48} is a macroscopic force arising from quantum-mechanical and thermal fluctuations of the electromagnetic field.  This force is the macroscopic version of the van der Waals force including finite light speed effects. Its properties qualitatively depend on the electric and magnetic susceptibilities of the materials involved in a broad frequency range \cite{Lifshitz55}. Hence, the discovery of new materials opens up a wealth of possible research avenues \cite{Woods2016Review}. 
 
 Since the fluctuations cannot be shut down in any way, Casimir interactions are ubiquitous. Therefore, the 
 Casimir force must always be included in the search for hypothetical new forces and gravitational tests in the nanometer to micron range \cite{Chen2016, Ren2026Arxiv}. Measuring the Casimir force with sufficient precision requires modern technology, such as micro-mechanical switches and actuators, whose actuation dynamics is affected by the Casimir force \cite{Munday2021Review, Broer2025}. The ubiquitous nature of Casimir interactions also leads to their occurrence in natural systems. Examples include spinning asteroids with small masses \cite{Rozitis2014} and the pre-melting of ice \cite{LiY2022}. Another recent development is the study of Casimir interactions in macro-biomolecular systems such as peptides \cite{Klimchitskaya2021, Klimchitskaya2026} or DNA \cite{Ge2023, Broer2026}. This may be the first step towards investigating Casimir interactions in biological systems, and it is relevant for potential applications in nanotechnology \cite{Bui2019}. 
 
 Water is crucial for every aspect of life on this planet. Remarkably, in nanometer-sized cavities its properties change completely compared to the bulk case \cite{LiQ2015WaterReview}. In  particular, in planar nanocavities, water turns out to be anisotropic: light propagates at different speeds through it, depending on the direction. This was experimentally demonstrated by measuring the perpendicular \cite{Fumagali2018} and parallel \cite{Wang2025} components of the dielectric function. Bulk water possesses no such asymmetry and is completely isotropic \cite{Gudarzi2021}. 
 
%  \note{Maybe some additional literature could be cited here.}

 The isolation of graphene \cite{Novoselov2004}, a two-dimensional allotrope of carbon and world's thinnest material, marked the beginning of two-dimensional physics. This material possesses many  remarkable properties, such as high electron mobility \cite{ZhangYB2009} and a large mechanical strength \cite{Booth2008}. A vast body of literature on graphene exists; see Ref. \onlinecite{Urade2023} for a recent review. 
 
 Investigations of the combination between the Casimir force and graphene started not long after graphene's initial isolation \cite{Bordag2006}. Since then, a remarkable property has been predicted: unlike conventional materials,  the Casimir force with graphene exhibits a sensitive temperature-dependence at submicron-scale separations \cite{Santos2009,Svetovoy2011}. Later, this effect was experimentally demonstrated \cite{LiuMPRL2021,LiuMPRB2021}. Other research that combines the Casimir force with graphene includes, but is not limited to: Multiple stacks of free-standing graphene \cite{Drosdoff2010},   an experiment with a silica substrate \cite{Banishev2013}, Casimir torque with anisotropic twisted bilayer graphene \cite{RodriguezLopez2023}, out-of-equilibrium Casimir forces \cite{Klimchitchkaya2025}, the impact of different graphene conductivity models \cite{RodriguezLopez2025}, graphene with drift currents \cite{KeM2026}, and the influence of the Casimir force on flexural modes of suspended graphene \cite{Ribeiro2026}.

 Since the experimental realization of the anomalously low static perpendicular component of the permittivity of water in planar nanocavities \cite{Fumagali2018}, its consequences for the macroscopic van der Waals force have been explored \cite{Esquivel2020}. However, it has turned out that the parallel component is even more strongly affected \cite{Wang2025}, and that dielectric function of nanoconfined water is significantly influenced by the properties of the surrounding materials \cite{Papadopoulou2021}. Moreover, for Casimir force calculations, the static permittivities are not enough, and the frequency-dependent, complex dielectric function is required. Recently, an molecular dynamics simulation was reported for a nanometer-thin water layer confined by graphene in the frequency range from GHz to $\sim$100 THz \cite{Becker2024}. Here, we combine this result with the measured permittivity components from Refs. \onlinecite{Fumagali2018,Wang2025} as an input for our Casimir force calculations for water layer of a thickness between 0.7 nm and 2 microns. In addition, we propose a simple, phenomenological model for the influence of the confined water on graphene. This way, we take into account the non-trivial influence of the highly anisotropic confined water on the Casimir force between the graphene sheets. It is worth noting that Casimir forces, with a few exceptions \cite{,Sarlah2001,Deng2015, Mostepanenko2015}, are typically considered for isotropic gaps.
 
 Graphene offers several advantages as a confining material. Firstly, it has no roughness, which, for conventional materials, tends to be of the nano-scale, comparable to the water layer thickness.  Another advantage is, that under ambient conditions and without edge effects, does not react chemically with water.

\begin{figure}[!hbt]
    \centering
    \includegraphics[width=0.45\textwidth]{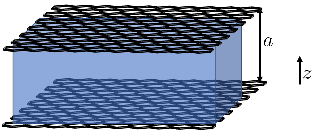}
    \caption{  A water layer confined by graphene sheets (not to scale). The water layer is modeled as an anisotropic 3D bulk medium with a thickness $a$ whereas the graphene sheets are treated as 2D. Outside the graphene sheets there is vacuum.  %\note{This picture is unremarkable}  
    }
    \label{fig:Geometry}
\end{figure}

 \section{Methodology}

 \subsection{Casimir Force}

To obtain the Casimir force, we must solve the Maxwell equations \cite{Lifshitz61}. One way to do this, is by treating the Maxwell equations as an eigenvalue problem, a method known as the transfer matrix method \cite{Berreman1972,Yeh1979}. The transfer matrix method can be applied in the context of Casimir interactions. This is particularly useful in the case of in-plane anisotropy, where the usual $s-$ (TE) and $p-$ (TM) mode decomposition fails (See for example, Refs. \onlinecite{Broer2019,RodriguezLopez2020,Broer2023}). 

For the water layer in \cref{fig:Geometry} anisotropy lies out-of-plane, and the dielectric tensor is given by

\begin{equation}
 \wm{\varepsilon}=\diag\left(\varepsilon_\parallel,\varepsilon_\parallel,\varepsilon_\perp\right).
\end{equation}
In this case,  $s-$ and $p-$ modes are eigenfunction solutions of the Maxwell equations. The eigenvalues, which represent the four possible values of the $z-$component of the wave vector in the anisotropic water layer in \cref{fig:Geometry}, are given by
 \begin{subequations}
 \begin{align}
 q^\pm_s=\pm\sqrt{\varepsilon_{\parallel}\tfrac{\omega^2}{c^2}-k_\rho^2}\\
 q^\pm_p = \pm\sqrt{\varepsilon_{\parallel}\tfrac{\omega^2}{c^2}-\tfrac{\varepsilon_{\parallel}}{\varepsilon_{\perp}}k_\rho^2},
 \end{align}
\end{subequations}
 where $\omega$ denotes the angular frequency and $c$ is the speed of light in vacuum. The positive and negative signs correspond to the forward and backward propagating modes, respectively, and $k_\rho=\sqrt{k_x^2+k_y^2}\geq0$ denotes the radial component of the wave vector, parallel to the surfaces.

This leads to the following expressions for the Fresnel reflection coefficients:

\begin{subequations}\label{eq:Fresnel}
 \begin{gather}
  r_s = \frac{q_s- k_0-Z_0\omega\sigma/c}{q_s + k_0+ Z_0 \omega\sigma/c} \\
  r_p = \frac{q_p-\varepsilon_\parallel k_0+ Z_0ck_0q_p\sigma/\omega}{q_p+\varepsilon_\parallel k_0+Z_0ck_0q_p\sigma/\omega} , 
 \end{gather}
\end{subequations}
where $\sigma$ denotes the conductivity in the local Kubo model, $Z_0=\mu_o c$ the vacuum impedance, and $k_0=\sqrt{\omega^2/c^2-k_\rho^2}$, the vacuum eigenvalue.
For $\sigma=0$, the known reflection coefficients for an anisotropic slab enclosed by two isotropic half-spaces (see e.g. Refs. \onlinecite{Greenaway1970, Sarlah2001, Deng2015, Mostepanenko2015}) are recovered. For the isotropic case, where $q_s=q_p$, see, for example, Ref. \onlinecite{RodriguezLopez2025}.

%\note{ factor for $\sigma$ in \cref{eq:Fresnel}. $s$ and $p$ modes mixed up?}

These coefficients signify how much of the EM field radiation of a given input polarization is reflected as an $s-$ or $p-$polarized wave. The reflected polarization is equal to the input, hence $r_{sp}$ and $r_{ps}$, are zero in this case. %The anisotropy lies out of the plane of reflection and the $s-$ and $p-$polarized modes are solutions throughout all media considered here.

 The Lifshitz formula for the Casimir free energy per unit area at temperature $T$ isotropic half-spaces separated by an uniaxial anistropic gap is \cite{Sarlah2001, Deng2015, Mostepanenko2015}:
 
 \begin{equation}\label{eq:LifshitzE}
 \begin{split}
    &\mathcal{E}_\text{Cas}(a)= \frac{k_b T }{2\pi}\sum_{n=0}^{\infty}(1-\tfrac{1}{2}\delta_{0n})\times\\
    &\int\limits_0^\infty k_\rho \ud k_\rho\Big(\ln[1-r_{p}(i\zeta_n,k_\rho)^2e^{-2q_p a}]+\\
    &\ln[1-r_{s}(i\zeta_n,k_\rho)^2e^{-2q_s a}]\Big),\\
    {}\\%trying to prevent equation from interfering with references
    {}
 \end{split}
 \end{equation}
where $k_b$ denotes the Boltzmann constant and all  quantities are evaluated at the imaginary Matsubara frequencies $\zeta_n=2\pi n k_b T/\hbar$, $n=0,1,2...$, with $\hbar$ the reduced Planck constant.

Since the anisotropy is out-of-plane, it does not break the rotation symmetry and the integral in \cref{eq:LifshitzE} can be reduced to a single variable. The corresponding Casimir force per unit area (or pressure) is then obtained from:

\begin{equation}
                 P_\text{Cas}(a)=-\frac{\partial \mathcal{E}_\text{Cas}(a)}{\partial a},
\end{equation}
More explicitly this is

\begin{equation}\label{eq:LifshitzP}
\begin{split}
 &P_\text{Cas}(a)=-\frac{k_b T}{\pi}\sum_{n=0}^{\infty}(1-\tfrac{1}{2}\delta_{0n})\times\\
 &\int\limits_0^\infty k_\rho \ud k_\rho\Bigg(\frac{q_pr_p(i\zeta_n,k_\rho)^2}{e^{2q_p a}-r_p(i\zeta_n,k_\rho)^2}+\\&\frac{q_s r_s(i\zeta_n,k_\rho)^2}{e^{2q_s a}-r_s(i\zeta_n,k_\rho)^2}\Bigg)
\end{split}
\end{equation}
Numerically, it is convenient to rescale to the dimensionless quantities: $\xi_n\equiv\zeta_n/\zeta_c$, with the characteristic frequency $\zeta_c\equiv c/(2a)$,  $x\equiv2aq_p\sqrt{\varepsilon_\perp/\varepsilon_\parallel}=2a\sqrt{\varepsilon_\perp\zeta_n^2/c^2+k_\rho^2}$, and $y\equiv2aq_s$. Since $k_\rho\geq0$ and $\varepsilon_{\parallel,\perp}\geq1$, the integration limits of the Lifshitz formula change \cite{Mostepanenko2015}:

\begin{equation}\label{eq:LifshitzPFinal}
\begin{split}
 P(a)=-\frac{k_b T}{8\pi a^3}\sum_{n=0}^{\infty}(1-\tfrac{1}{2}\delta_{0n})\times\\
\Bigg[\sqrt{\frac{\varepsilon_\parallel}{\varepsilon_\perp}}\int\limits_{\xi_n\sqrt{\varepsilon_\perp}}^\infty \frac{x^2r_p^2\exp\left(-x\sqrt{\varepsilon_\parallel/\varepsilon_\perp}\right)}{1-r_p^2\exp\left(-x\sqrt{\varepsilon_\parallel/\varepsilon_\perp}\right)}\ud x\\
+\int\limits_{\xi_n\sqrt{\varepsilon_\parallel}}^\infty\frac{y^2r_s^2\exp(-y)}{1-r_s^2\exp(-y)}\ud y\Bigg],  
\end{split}
\end{equation}
where the arguments of $\varepsilon_{\parallel,\perp}$ and $r_{p,s}$ have been suppressed. The reflection coefficients change to

\begin{subequations}
\begin{align}
 r_s=\frac{y-\sqrt{y^2-(\varepsilon_\parallel-1)\xi_n^2}-\Sigma_s}{y+\sqrt{y^2-(\varepsilon_\parallel-1)\xi_n^2}+\Sigma_s}\\
 r_p=\frac{x\sqrt{\varepsilon_\parallel/\varepsilon_\perp}-\varepsilon_\parallel\sqrt{x^2-(\varepsilon_\perp-1)\xi_n^2}+\Sigma_p}{x\sqrt{\varepsilon_\parallel/\varepsilon_\perp}+\varepsilon_\parallel\sqrt{x^2-(\varepsilon_\perp-1)\xi_n^2}+\Sigma_p},
\end{align}
\end{subequations}
where

\begin{subequations}\label{eq:Sigma}
\begin{align}
\Sigma_s\equiv\xi_n Z_0\sigma(i\zeta)\\
\Sigma_p\equiv \sqrt{\varepsilon_\parallel/\varepsilon_\perp}Z_0\sigma(i\zeta) \frac{2ak_0 x}{\xi_n}.
\end{align}
\end{subequations}

%The  Fresnel reflection coefficients are given by \cref{eq:Fresnel}.

 \subsection{Material properties}

 \subsubsection{The conductivity of the graphene sheets}\label{sec:model}

  % \note{ Is full QFT treatment necessary? Need to present a good argument to justify this choice. Here is my attempt:}
 
 %how the interplay between the dielectric functions of the water layer and the confining graphene sheets affect the Casimir force.
 
 We are primarily interested in the influence of water confinement on the Casimir force. Therefore, our aim is not necessarily to present the most realistic calculation, but rather - as a proof of concept - describe this in terms of simple models. Moreover, the difference between Casimir forces  based on the Dirac model and the Kubo model appears to be very small ($<0.1\%$) in the case of a vacuum gap \cite{RodriguezLopez2025}.
 
%  \begin{figure}[hb]
%     \centering
%     \includegraphics[width=0.45\textwidth]{KuboParameters}
%     \caption{  The parameters of the Kubo model compared to the bulk case, where there is no thickness dependence. Here $E_{F\text{bulk}}=0.15$ eV and $\Gamma_\text{bulk}=0.91$ THz. }
%     \label{fig:KuboParam}
% \end{figure}
%  

  Let us start with the simplest possible model for the conductivity of graphene.  This is the local Kubo model, developed in Ref. \onlinecite{Falkovsky2007}.  By 'local' it is meant that spatial dispersion is ignored. In this limit where $k\rightarrow0$, the conductivity is a sum of inter- and intraband contributions: $\sigma(\omega)=\sigma_\text{intra}+\sigma_\text{inter}$, where \cite{Falkovsky2008,RodriguezLopez2025Conductivity}
  
% \begin{eqnarray}\label{eq:sigma1}
%  &\sigma_\text{intra}(\omega) = \frac{1}{\pi\hbar}\frac{2ik_b Te^2/\hbar }{\omega + i\Gamma}\ln\left(2\cosh\left(\frac{E_F}{2 k_b T}\right)\right),\\\nonumber 
%       &\sigma_\text{inter}(\omega)  = \frac{ e^2/\hbar }{4}G\left(\left|\frac{\hbar\omega}{2}\right|\right)
%       +\\\nonumber
%       &\frac{ie^2/\hbar}{4}\frac{4\hbar\omega}{\pi}\int\limits_0^\infty  \ud\xi \frac{G(\xi)-G\left(\left|\frac{\hbar\omega}{2}\right|\right)}{(\hbar\omega)^2-4\xi^2},
%  \end{eqnarray}
 
\begin{subequations}\label{eq:sigma1}
 \begin{align}
   &\sigma_\text{intra}(\omega) = \frac{1}{\pi\hbar}\frac{2ik_b Te^2/\hbar }{\omega + i\Gamma}\ln\left(2\cosh\left(\frac{E_F}{2 k_b T}\right)\right),\\\nonumber 
      &\sigma_\text{inter}(\omega)  = \frac{ e^2/\hbar }{4}G\left(\left|\frac{\hbar\omega}{2}\right|\right)
      +\\
      &\frac{ie^2/\hbar}{4}\frac{4\hbar\omega}{\pi}\int\limits_0^\infty  \ud\xi \frac{G(\xi)-G\left(\left|\frac{\hbar\omega}{2}\right|\right)}{(\hbar\omega)^2-4\xi^2},
 \end{align}

\end{subequations}

with the shorthand notation 

\begin{equation}
 G(E)=\frac{\sinh(\beta E)}{\cosh(\beta E_F)+\cosh(\beta E)}.
\end{equation}
Here, $e$ denotes the electron charge, $\beta=1/(k_b T)$, $\Gamma$ the relaxation frequency of 0.91 THz, \cite{LiZQ2008} and $E_F$ the Fermi energy. Here we assume that the band gap is zero, because graphene retains its semi-metal properties in the presence of the water. Moreover, we use the Fermi energy as synonymous with the chemical potential, neglecting the temperature effect on this parameter.  We note that the charge density was measured to be \cite{Schedin2007} $n_s\approx1.5\cdot10^{12}$ cm$^{-2}$, and due to the linear dispersion of graphene, we have  

\begin{equation}\label{eq:EFBulk}
 E_F=\hbar v_F\sqrt{\pi n_s}\approx0.14 \quad\text{eV},
\end{equation}                                                                                                                                                           
where $v_f\approx c/300$ is the Fermi velocity.

At imaginary frequencies \cref{eq:sigma1}, becomes

\begin{subequations}
 \begin{align}
    &\sigma_\text{intra}(i\zeta) = \frac{2k_bTe^2/\hbar}{\pi\hbar(\zeta+\Gamma)}\ln\left(2\cosh\left(\frac{E_F}{2 k_b T}\right)\right),\\
    &\sigma_\text{inter}(i\zeta) = \frac{e^2}{4\hbar}G\left(\frac{\hbar\zeta}{2}\right)+\frac{e^2}{4\hbar}\frac{4\hbar\zeta}{\pi}\int\limits_0^\infty\frac{G(\xi)-G(\hbar\zeta/2)}{\hbar^2\zeta^2+4\xi^2}\ud\xi.
 \end{align}
\end{subequations}

\begin{figure*}[tp]
    \centering
    \includegraphics[width=0.8\textwidth]{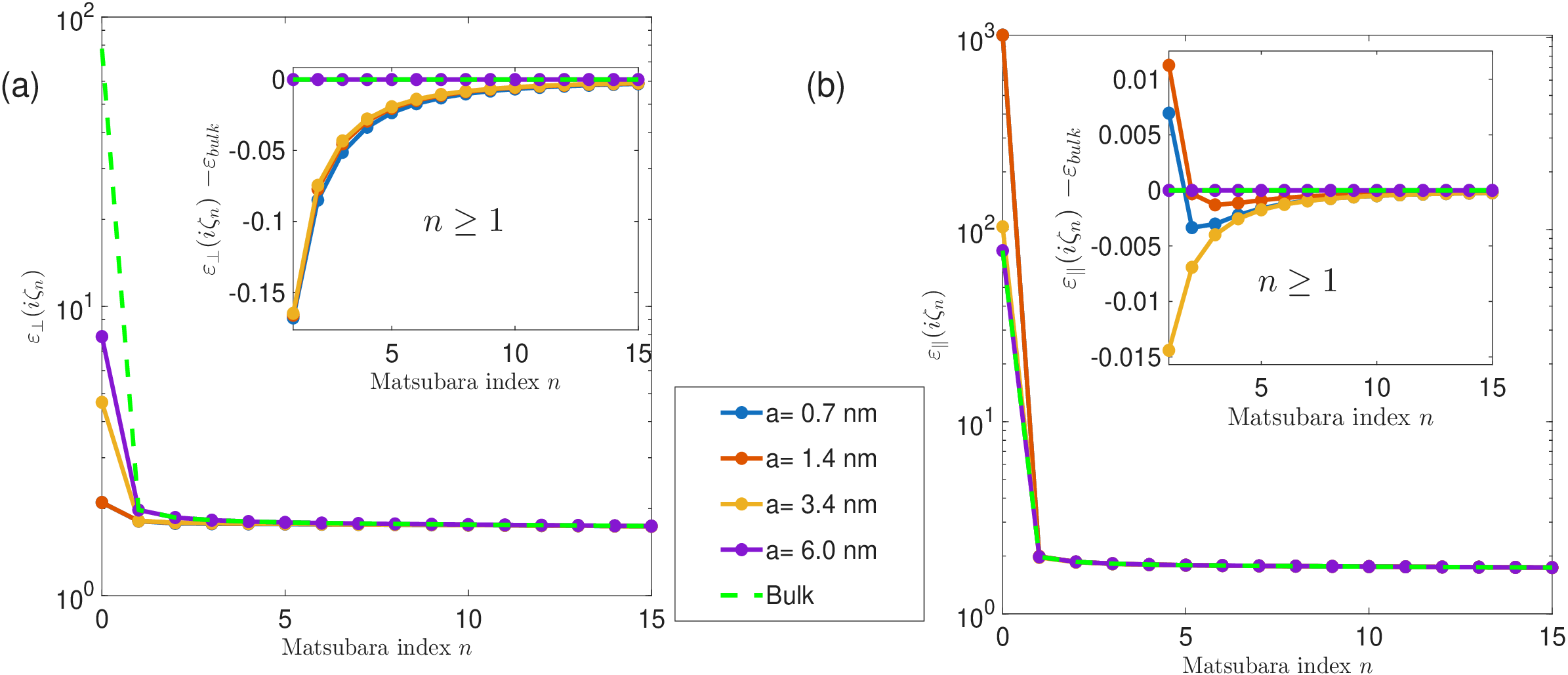}
    \caption{ The permittivity components of confined water at different thicknesses. Panel (a):  
    the perpendicular component
    Panel (b): the parallel component. Insets: deviations from the bulk case. The perpendicular and parallel  components for $n=0$ were taken from Refs. \onlinecite{Fumagali2018,Wang2025}, respectively. The  components for $n>0$  were taken from Ref. \onlinecite{Becker2024}.
    The bulk permittivity was taken from Ref. \onlinecite{Gudarzi2021}. %\note{figure placement to be fixed here}
    } 
    \label{fig:epsIZeta}
\end{figure*}

 \subsubsection{The permittivity of the water layer}

 Let us start with the static permittivity components reported in Refs. \onlinecite{Fumagali2018,Wang2025}. The perpendicular component is given by\cite{Fumagali2018}
 
 \begin{equation}\label{eq:eps_perp_stat}
  \varepsilon_\perp(0,a) = 
 \begin{cases}
    \min(\varepsilon_\perp)\quad a<2d_i\\
   \frac{a}{2d_i/\min(\varepsilon_\perp)+(a-2d_i)/\varepsilon_\text{bulk}(0)}\quad a\geq2d_i,
 \end{cases}
 \end{equation}
where $d_i=0.75$ nm, $\varepsilon_\text{bulk}(0)\approx78$, and $\min(\varepsilon_\perp)\approx2.1$. 
\cref{eq:eps_perp_stat} models the perpendicular component as a double capacitor with two 'dead layers' each thickness $d_i$. However, the parallel component cannot be modeled in the same way, because it decreases much faster as a function of the thickness. \cite{Wang2025} Here, we fit it to an exponential curve:

\begin{equation}\label{eq:eps_par_stat}
\varepsilon_\parallel(0,a)\approx
\begin{cases}
    \max(\varepsilon_\parallel)\quad a<2d_i\\
  \max(\varepsilon_\parallel)\exp\left(-\frac{a-2d_i}{2d_i}\right)+\varepsilon_\text{bulk}\quad a\geq2d_i.
\end{cases}
\end{equation}

 %Their thickness dependences are given by \cref{eq:eps_perp_stat,eq:eps_par_stat}, respectively. 
 We will use these permittivity components as inputs for the $n=0$ contribution to the Matsubara sum \cref{eq:LifshitzP}. We stress that at $a=0.7$ nm, the permittivity components were not measured and we used the extrapolated values of $\varepsilon_\parallel\approx1000$ and $\varepsilon_\perp\approx2.1$.

For the frequency range between 39 and 118 THz, that is, for Matsubara indices $1\leq n\leq3$,  we use the effective complex, frequency-dependent susceptibility components $\chi^\text{eff}_{\perp,\parallel}(\omega)$ reported in Ref. \onlinecite{Becker2024}. The constitutive relation is simply

\begin{equation}
 \varepsilon_{\parallel,\perp}(\omega,a) = 1+\chi^\text{eff}_{\parallel,\perp}(\omega,a),
\end{equation}
which, for the Casimir force calculations, can be transformed to imaginary frequencies via the Kramers-Kronig relation:

\begin{equation}\label{eq:KK}
 \varepsilon_{\parallel,\perp}(i\zeta,a)=1+\frac{2}{\pi}\int_0^{\infty}\frac{\text{Im}(\chi^\text{eff}_{\parallel,\perp}(\omega,a))\omega\ud\omega}{\zeta^2+\omega^2}
\end{equation}
where Im$(\chi^\text{eff}_{\parallel,\perp}(\omega,a))\geq0$, because it represents an absorption spectrum. Hence, \cref{eq:KK} ensures that $\varepsilon_{\parallel,\perp}$ is positive and monotonically decreasing as a function of $\zeta$. The purpose of this transformation is to facilitate numerical calculations.

At sufficiently high frequencies, the photons do not interact with the motions of the water molecules in the gap, because they operate on a different timescale. Effectively, this means that the permittivity becomes its bulk value:

\begin{equation}
 \varepsilon_{\parallel,\perp}(i\zeta_n,a)\rightarrow\varepsilon_\text{bulk}(i\zeta_n)\quad n>3.
\end{equation}

For the bulk case, which is isotropic and independent of the thickness, we use the permittivity reported in Ref. \onlinecite{Gudarzi2021}, which is given by

\begin{equation}\label{eq:epsBulk}
    \varepsilon^{GA}_\text{bulk}(i\zeta) = 1+\sum\limits_{j=1}^3\frac{C_j}{1+(\zeta/\omega_j)^{\beta_j}},
\end{equation}
whose parameters are $C_1=73.48$, $\hbar\omega_1=8.1\cdot10^{-5}$ eV, $\beta_1=0.988$, $C_2=2.534$, $\hbar\omega_2=0.016$ eV, $\beta_2=1.1$, $C_3=0.755$, $\hbar\omega_3=16.1$ eV, and $\beta_3=1.751$.

 To properly describe the transition from the confined to the bulk case, we must compare the bulk permittivities according to Refs. \onlinecite{Gudarzi2021} and \onlinecite{Becker2024}.  It turns out they are not identical; we denote the difference by $\Delta$:
 
 \begin{equation}\label{eq:deficit}
  \Delta(i\zeta) \equiv \varepsilon^{GA}_\text{bulk}(i\zeta)-\varepsilon^{BN}_\text{bulk}(i\zeta),
 \end{equation}
which tends to $0.75\approx C_3$ for $n\ll100$. Strictly speaking, the deficit $\Delta$ depends on frequency, but it turns out to be almost constant for $n<3$, with a value close to $C_3$ in \cref{eq:epsBulk}.  This is not a coincidence: $\hbar\omega_3$ corresponds to $\sim3.9\cdot10^{3}$ THz, which is much higher than 125 THz, largest frequency considered in Ref. \onlinecite{Becker2024}, which is why it appears to be constant at the at Matsubara indices $n\ll100$. The term for $j=3$ in \cref{eq:epsBulk} represents the electron oscillations, which are not included in the Born-Oppenheimer molecular dynamics simulation of Ref. \onlinecite{Becker2024}. Rather, the simulations include the motion of the atomic nuclei in the confined water: reorientation (Debye $\sim15$ GHz), the H bond stretch ($\sim$ 5 THz), libration ($\sim20$ THz), the HOH bend ($\sim50$ THz), and the OH stretch ($\sim$ 100 THz). At higher frequencies, no such motion exists.  The nearly constant, real offset $\Delta$ is invisible to the Kramers-Kronig transform of \cref{eq:KK}; adding it will not affect compliance with it in the $\sim$1 to $\sim$100 THz range.
 
   \begin{figure*}[!tp]
    \centering
    \includegraphics[width=0.8\textwidth]{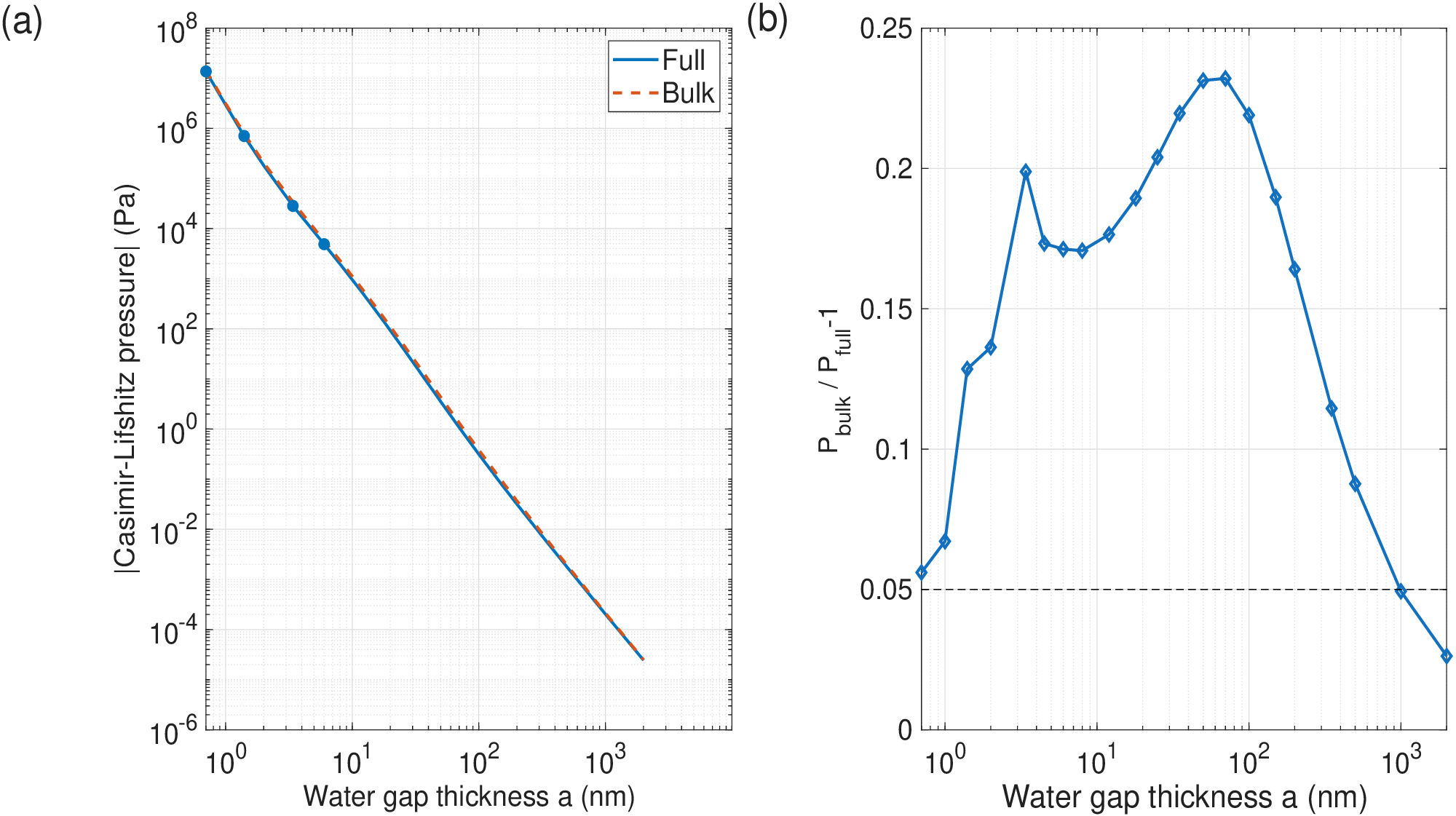}
    \caption{ (a) The Casimir pressure. The blue circles denote thicknesses sampled in Ref. \onlinecite{Becker2024}.  (b) The relative effect of confinement on the Casimir pressure. The dashed line shows where this is less than 5\%. %\note{figure placement to be fixed here} 
    }
    \label{fig:FullVsBulk}
\end{figure*}

Hence, we need to add this offset to effective susceptibility data from Ref. \onlinecite{Becker2024}:

\begin{equation}\label{eq:epsFinal}
\varepsilon_{\parallel,\perp}(i\zeta,a) = 1+ \chi^\text{eff}_{\parallel,\perp}(i\zeta,a)+\Delta(i\zeta).
\end{equation}
These quantities are plotted in \cref{fig:epsIZeta}. It can be seen that, with the addition of the term $\Delta$, all permittivity components decrease monotonically with $\zeta$, and that the perpendicular component converges to the bulk case at around $n\sim10$, whereas for the parallel component, this happens immediately at $n=1$. We now have enough information to proceed with the Casimir force calculation.

\section{Results and Discussion}

%  Firstly, we must distinguish between confinement and interface effects. The former describes the effect of the graphene walls imposing restrictions on the movement of the water molecules in the gap. This is strictly associated with very small thicknesses $a<6$ nm. \cite{Becker2024} The latter effect is due to the proximity of some of the water molecules to the confining surface. This can persist at much larger thicknesses, of up to $a=100$ nm or even more \cite{Fumagali2018}. \cref{eq:eps_perp_stat} shows slow convergence to bulk case, due to interface effect: the ``dead layer'' of water molecules close to the graphene surfaces will become negligible only at distances much greater than $2d_i=1.5$ nm.

 Typically in Lifshitz theory calculations for aqueous systems, the $n=0$ term is considered dominant. \cite{Parsegian1969, Ninham1970}  However, due to screening by mobile ions beyond the Debye length this dominance diminishes. \cite{Davies1972, Netz2001} The Debye length is $\sim1\mu$m for ideally pure water, but it can be smaller in more realistic situations ($\sim 100$ nm).

Throughout this paper, we let $T=$300 K. For studies that explore the sensitivity of the temperature-dependence of the Casimir force with graphene, see, for example Refs. \onlinecite{Santos2009,Svetovoy2011,LiuMPRL2021,LiuMPRB2021}.

So far, we have neglected the effect of the modification of the graphene conductivity under the influence of water confinement. %the model in Section \ref{sec:model} is only applicable to the $n=0$ Matsubara term. 
This will have no effect on the Casimir force. To understand this, we examine \cref{eq:Fresnel,eq:Sigma}. It can be seen that, at $n=0$, $r_p=1$ and $r_s=0$. Therefore, the exact values of the conductivity do not affect the Casimir force. This is almost certainly an artifact of the somewhat simplistic model introduced here. A new, more general model will have to be developed to rectify this.

The results of the implementation of \cref{eq:LifshitzPFinal} are shown in \cref{fig:FullVsBulk} (a). We stress that, at short ranges ($a<6$ nm), we currently have information about only three data points, which were sampled in Ref. \onlinecite{Becker2024}. The Casimir pressure is attractive (negative) because the configuration in \cref{fig:Geometry} is symmetric. A repulsive Casimir force between non-magnetic semi-infinite plates requires two different surrounding materials such that the permittivity of the intervening liquid is larger than one of the surrounding materials, but smaller than the other one.  \cite{Lifshitz61, Deng2015}

 \begin{figure}[!htpb]
    \centering
    \includegraphics[width=0.49\textwidth]{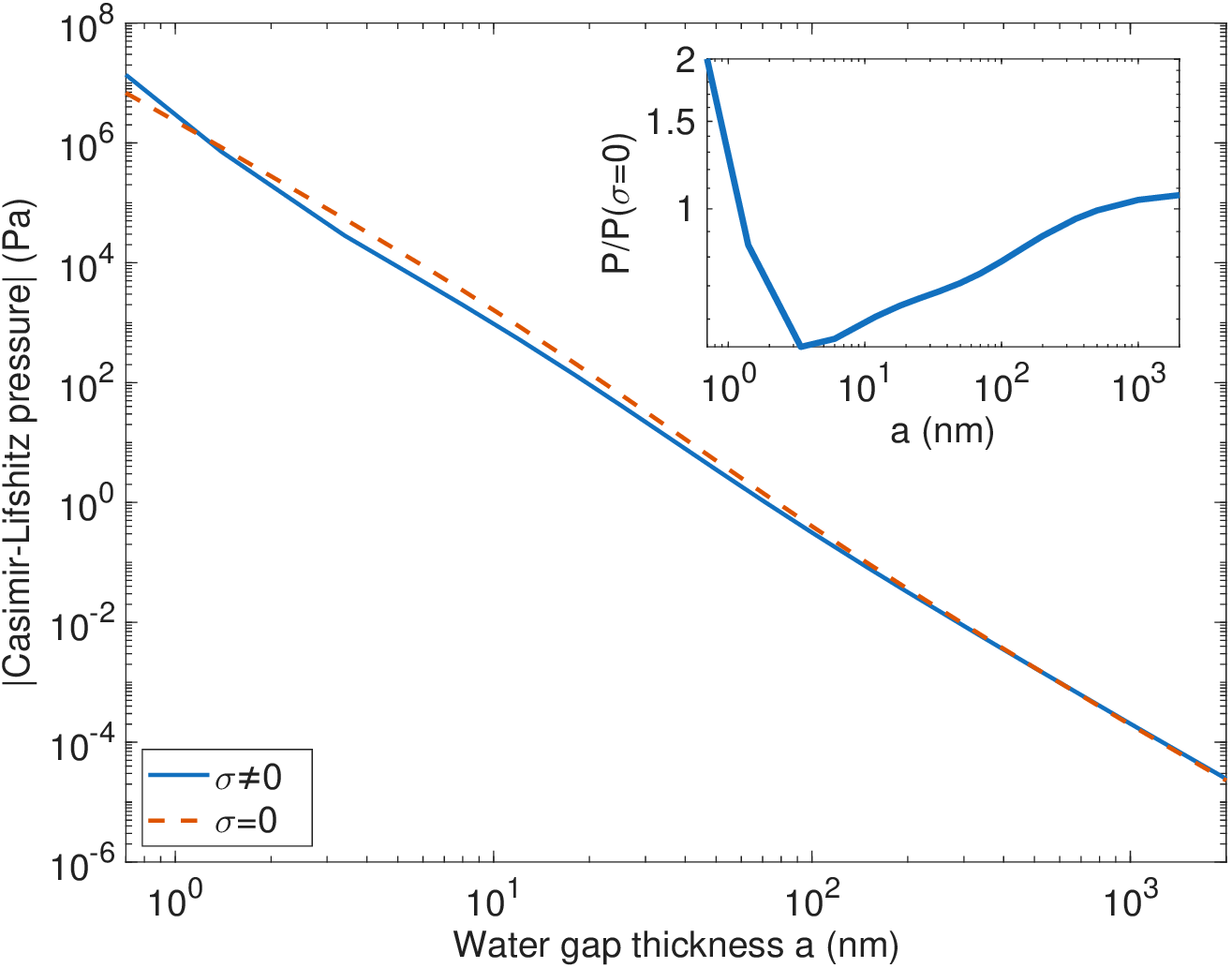}
    \caption{The role of graphene. The solid line represents the full Casimir pressure, and the dashed line the case where graphene is 'turned off'  ($\sigma=0$). The inset shows the ratio between these cases. }
    \label{fig:RoleOfGraphene}
\end{figure}

To illustrate the role of the graphene, we have compared the total Casimir pressure to the case where $\sigma=0$ in \cref{fig:RoleOfGraphene}. In the inset, it can be seen that the ratio between these cases rapidly decreases until $a=3.4$ nm, after which it slowly increases to a near constant value, where it is close to unity. We stress that this is the result for a free-standing water film confined by graphene sheets according to the local Kubo model. A substrate of e.g. silica or a non-local conductivity model could change this significantly.

\begin{figure}[!htpb]
    \centering
    \includegraphics[width=0.49\textwidth]{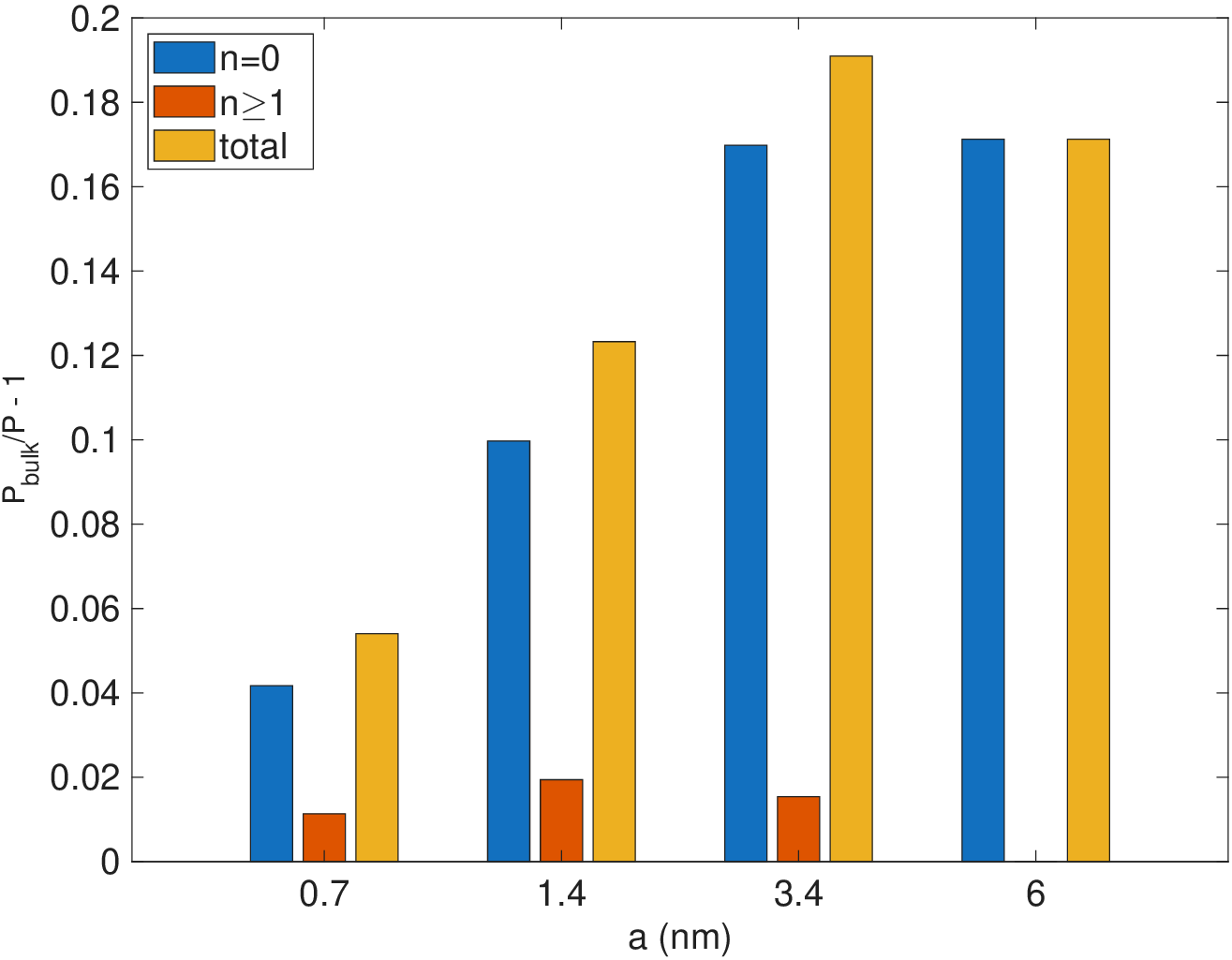}
    \caption{ Decomposition of the Matsubara sum at small thicknesses $a\leq6$ nm.}
    \label{fig:MatsubaraDecomp}
\end{figure}

 In \cref{fig:FullVsBulk} (b) it can be seen that the bulk model does not produce the correct Casimir force. At short range ($a<6$ nm), it overestimates the Casimir force by up to 20 \%. This is mostly due to the static permittivity components ( the $n=0$ term. It can be seen \cref{fig:MatsubaraDecomp} that the $n\geq1$ contribution is only few percent of the total.

 %This can only to a small extent be  attributed to the  parallel component ($\varepsilon_\parallel\approx1000$) at $n=0$, which constitutes only a tiny fraction ($<0.01\%$) of the total Casimir force. The vast majority of the confinement effect is due to $n>0$ terms, which combine the data of Ref. \onlinecite{Becker2024} for confined water with the bulk dielectric function of Ref. \onlinecite{Gudarzi2021}.
 
 At thicknesses of 6 nm or larger, we modeled the perpendicular component dielectric function as $\varepsilon_\perp(0,a)$ via \cref{eq:eps_perp_stat}, and as the bulk \cref{eq:epsBulk} for the parallel component and $n\neq0$.  This decreases monotonically, see \cref{fig:epsIZeta} (a). In \cref{fig:FullVsBulk} (b) we see that this effect can be as large as 24 \%. We must stress here that this is mainly due to our choice of the series capacitors 
  model \cref{eq:eps_perp_stat}. To our knowledge, there is no experimental research to compare this result to.
 
 Remarkably, at $a>6$ nm, this effect doesn't decrease monotonically. In \cref{fig:FullVsBulk} (b), it can be seen that this reaches a maximum around $a=70$ nm, after which it slowly decreases to the bulk limit. Only at $a=1$ $\mu$m does it reach a value within 5 \% of the bulk limit. This is mostly due to our choice of model \cref{eq:eps_perp_stat}, where the perpendicular component tends to its bulk value algebraically, rather than exponentially as in \cref{eq:eps_par_stat}. We repeat that this still needs to be checked experimentally. %However, this does not fully explain the non-monotonic behavior shown in \cref{fig:FullVsBulk} (b).

 To understand the non-monotonic behavior of anisotropy effect at $a>6$ nm, it is useful to realize that, all anisotropy effects come from the $n=0$ contribution. In the local Kubo model used here $r_p=1$, $r_s=0$ so that the Casimir pressure for $n=0$ becomes
 
 \begin{equation}
    P_0(a)=-\frac{k_b T\zeta(3)}{8\pi a^3}\frac{\varepsilon_\perp(0)}{\varepsilon_\parallel(0)}, 
 \end{equation}
where $\zeta(3)\approx1.202$ is the Ap\'ery constant.  
 This leads to the following relation

 \begin{equation}
\frac{P_\text{bulk}}{P_\text{full}}-1=\frac{f\delta}{1-f\delta},  
 \end{equation}
where 

\begin{equation}
f\equiv\frac{P_{0,\text{bulk}}}{P_\text{bulk}},  
\end{equation}
and $\delta$ represents the relative anisotropy:

\begin{equation}
 \delta\equiv1-\frac{\varepsilon_\perp(0)}{\varepsilon_\parallel(0)}. 
\end{equation}

\begin{figure}[!htpb]
    \centering
    \includegraphics[width=0.49\textwidth]{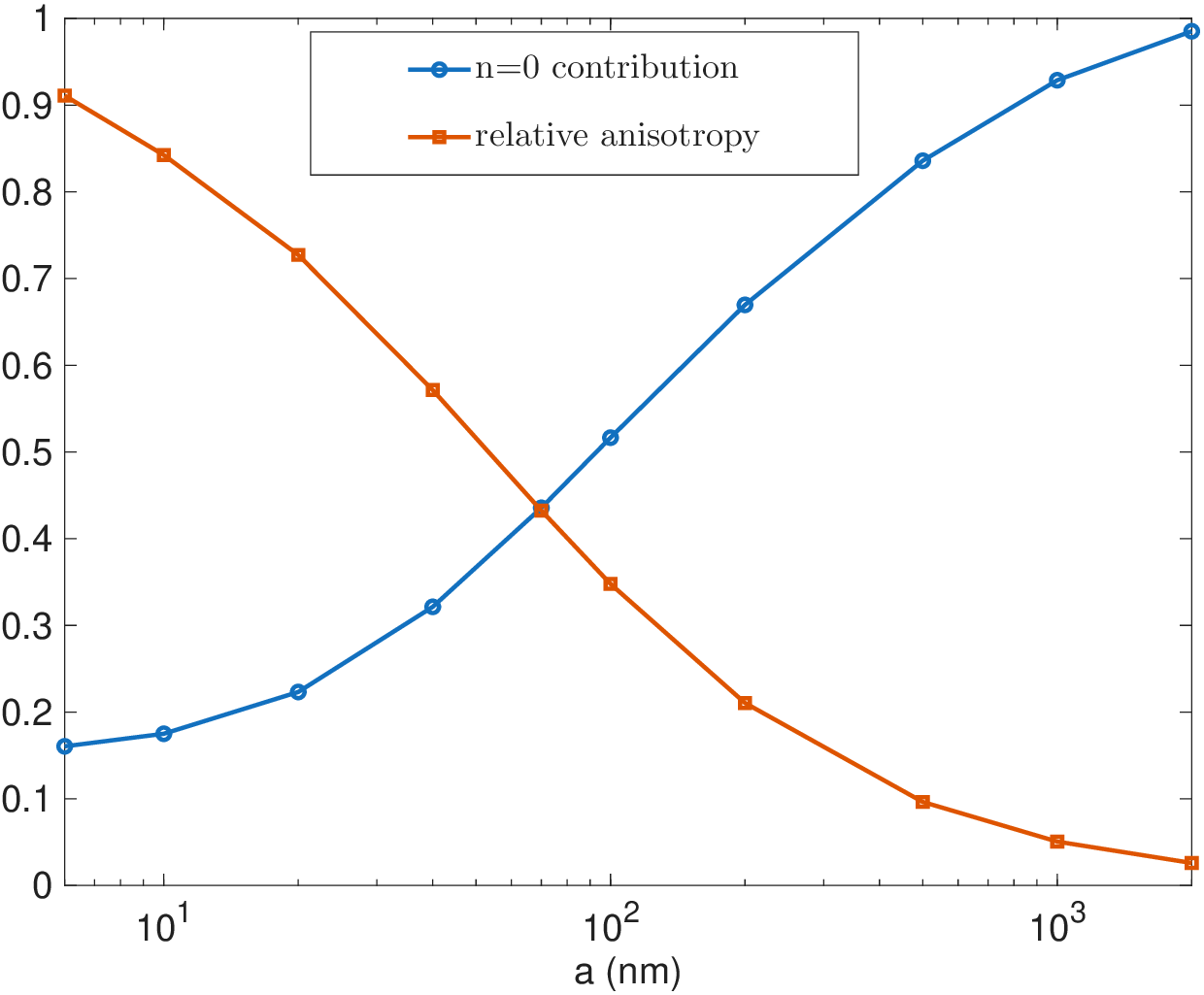}
    \caption{ Two competing effects in the range $a>$ 6 nm. The relative anisotropy $1-\varepsilon_\perp(0)/\varepsilon_\parallel(0)$ of the static dielectric constant decreases with the thickness $a$, whereas the $n=0$ Matsubara term becomes more dominant. }
    \label{fig:CompetingEffects}
\end{figure}

 We must therefore consider two competing effects, illustrated in \cref{fig:CompetingEffects}. 
 Any deviation from the bulk case purely originates from the $n=0$ contribution to the Matsubara sum, and the anisotropy is also associated with the same term. 
 While the relative anisotropy decreases with the thickness $a$, the relative contribution of $n=0$ term increases. Hence, on the one hand the water becomes more isotropic as a function of thickness, and on the other hand, the anisotropic $n=0$ term matters more. Therefore, the anisotropy effect due water molecules close to the graphene wall reaches a maximum around $a=70$ nm.

% \note{Compare to:  the non-retarded limit.}
% 
% It can be instructive to compare to the non-retarded limit, where $r_s=0$. 

\section{Conclusions and outlook}

 In summary, we investigated the effect of water confinement on the Casimir force by studying the Casimir force between two graphene sheets confining a nanometer-thin layer of water. The deviations from the bulk case were found to be up to 24\%. This may be due to our model choice of, ``dead layer'' double capacitor model of  \cref{eq:eps_perp_stat}.

 We hope that this paper will instigate more investigations on Casimir interactions with confined water. More information is needed about the full thickness- and frequency-dependence of the permittivity components of nano-confined water. We have assumed that the electron oscillations (the term for $j=3$ in \cref{eq:epsBulk}) are not affected by the graphene but this needs computational and experimental verification.  

 The local Kubo model in section \ref{sec:model} cannot describe the influence of the confined water on the graphene conductivity. Therefore a more general model is needed. Only then can this effect truly be assessed.

In the future, the role of the Casimir force in the premelting and formation of ice could be further investigated, \cite{LiY2022} including recent knowledge about the permittivities of nano-confined water. Perhaps this could shed some new light on the age-old question of the slipperiness of ice. \cite{Atila2025}
 
%\pagebreak

\begin{acknowledgments}
 We gratefully acknowledge R. Netz and M. Becker for the data of Ref. \onlinecite{Becker2024}. %\note{Funding info}
\end{acknowledgments}

%\pagebreak

\bibliography{Casimir3}
\end{document}